\documentclass[twocolumn,prl,showpacs]{revtex4}
\usepackage{graphicx}
\usepackage{epsfig}
\usepackage{bm}

\begin{document}

\title{
Dispersion Coefficients by a 
Field-Theoretic Renormalization of Fluid Mechanics
}

\author{Michael W. Deem$^1$}
\author{Jeong-Man Park$^{1,2}$}

\affiliation{
\hbox{}$^1$Chemical Engineering Department, University of
California, Los Angeles, CA  90095--1592}

\affiliation{
\hbox{}$^2$Department of Physics,
 The Catholic University of Korea, Puchon 420--743, Korea}

\begin{abstract}
We consider subtle correlations in 
the scattering of fluid by randomly placed obstacles,
which have been suggested to lead to a diverging
dispersion coefficient at long times for high 
P{\'e}clet numbers, in contrast to finite mean-field
predictions.
We develop a new master equation description of the fluid
mechanics that incorporates the physically relevant fluctuations, and
we treat those fluctuations by a renormalization group procedure.
We find a finite dispersion coefficient at low 
volume fraction of disorder and high P{\'e}clet numbers.
\end{abstract}


\maketitle

%47.55.Mh Flows through porous media
%05.60.+w Transport processes: theory

Dispersion of fluid particles by flow through random media is
important in problems ranging from pollutant transport 
through soil to enhanced oil recovery to chemical reactor design.
Recently, it has been suggested that the dispersion coefficient
may diverge logarithmically at long times \cite{Frenkel}.
Such a result, which is not inconsistent with experimental data on
flow through packed beds, would have significant implications.
This prediction, however, is in disagreement with mean-field
results that predict a finite dispersion coefficient \cite{Brady}.
Accounting for fluctuations with a renormalization group approach,
we find the dispersion coefficient should be finite.

There are four sources of randomness not accounted for
in the mean-field treatment of fluid dispersion.
The numerical results suggest a logarithmic divergence in
three dimensions, \emph{i.e.}\ that the upper critical dimension
may be three, and the mean-field theory, which lacks the randomness, 
would not be expected to detect such a divergence.
The first source of 
randomness is the random motion of the fluid particles.  Since 
the Navier-Stokes equations are the starting point for the mean-field theory,
all microscopic fluctuations of the fluid particles have
been suppressed. A second source of randomness is the
random locations of the obstacles to the fluid flow.
The mean-field theory is essentially a renormalized single-obstacle
theory.  As such, all correlations around an obstacle beyond the
Brinkman screening length are ignored.  The third source of
randomness present in the numerical simulations, but not present in the
mean-field theory, are mesoscopic fluctuations naturally present in the
lattice Boltzmann simulation method.  Imperfections in the pseudo-random
number generator are a fourth and final source of
randomness present in the simulations.
Using a field-theoretic renormalization group approach, we
take into account the first three physical sources of randomness.
We find that these sources of randomness do not modify the mean-field
prediction: the dispersion coefficient remains finite at long times and
high P{\'e}clet numbers.
The P{\'e}clet number is given by 
$Pe = v_x R / D_{\rm f}$, where $v_x$ is the average velocity of the
fluid, $R$ is the average radius of the blocking obstacles,
and $D_{\rm f}$ is the diffusion coefficient of the
pure fluid.

Our goal is to write a field theoretic representation of the
Navier-Stokes equation:
\begin{equation}
\partial_t v_i + 
    \textstyle{\prod_{ik}} \sum_j v_j \partial_j v_k
= \nu \nabla^2 v_i + f_i
\label{1}
\end{equation}
where $\nu = \mu/\rho$ is the kinematic viscosity, and 
$f_i = \Pi_{ik} (F_k - \partial_k P)/\rho$ is the
 total body force on the fluid.
The presence of the projection operator
 $\hat \Pi_{ik}({\bf k}) = \delta_{ik} - k_i k_k/k^2$
in these formulas ensures that the incompressibility condition 
$\nabla \cdot {\bf v}=0$
is maintained \cite{Forster}.  The Fourier transform is defined by
$\hat f({\bf k}) = \int d {\bf x} f({\bf x}) \exp(i {\bf k} \cdot {\bf x})$.

We first write a master equation model of fluid mechanics. 
Generalizing previous treatments from one dimension 
\cite{Petruccione} 
to $d=3$ dimensions,
we consider the fluid on 
a lattice, with ${\bf N}_{\mbox{\boldmath \scriptsize $\lambda$}}$ 
``velocity'' particles on lattice site ${\mbox{\boldmath $\lambda$}}$.
  Since the 
velocity is a vector, we denote the number of particles 
contributing to the $i^{\rm th}$ component of the 
velocity as $N_{\mbox{\boldmath  \scriptsize $\lambda$}}^i$.
The state of the system is defined by the configuration of
velocities in the system, and the master equation relates how
the probability of any given 
configuration, $P(\{ N_{\mbox{\boldmath  \scriptsize $\lambda$}}^i\} )$, 
changes in time
\begin{equation}
\partial_t P = A P
\label{2}
\end{equation}
where $A$ is a linear operator that specifies the dynamics of the fluid.
Defining  the identity, $\hat I$, shift operators, 
$\hat E_{\mbox{\boldmath \scriptsize $\lambda$}}^i 
 P(N_{\mbox{\boldmath \scriptsize $\lambda$}}^i) = 
P(N_{\mbox{\boldmath \scriptsize $\lambda$}}^i+1)$,
$\hat E_{\mbox{\boldmath \scriptsize $\lambda$}}^{i^{-1}}
 P(N_{\mbox{\boldmath \scriptsize $\lambda$}}^i) = 
P(N_{\mbox{\boldmath \scriptsize $\lambda$}}^i-1)$,
and a number operator,
$\hat N_{\mbox{\boldmath \scriptsize $\lambda$}}^i 
   P(N_{\mbox{\boldmath \scriptsize $\lambda$}}^i) 
= N_{\mbox{\boldmath \scriptsize $\lambda$}}^i 
 P(N_{\mbox{\boldmath \scriptsize $\lambda$}}^i)$,
the dynamics of Eq.\ \ref{2} are given by
diffusion, convection, and force terms,
 $A = A_{\rm d} + A_{\rm c} + A_{\rm f}$:
\begin{eqnarray}
A_{\rm d} &=& 
\frac{\nu } {h^2} \sum_{\mbox{\boldmath \scriptsize $\lambda$}} \sum_{jk}
\bigg[ 
\left(  \hat E_{{\mbox{\boldmath \scriptsize $\lambda$}} -{\bf e}_k}^{j^{-1}}
          \hat E_{\mbox{\boldmath \scriptsize $\lambda$}}^j - \hat I \right)
\nonumber \\
&&~~~~~~~~~+
\left( \hat E_{{\mbox{\boldmath \scriptsize $\lambda$}} +
   {\bf e}_k}^{j^{-1}} \hat E_{\mbox{\boldmath \scriptsize $\lambda$}}^j 
    - \hat I \right)
\bigg]
\hat N_{\mbox{\boldmath \scriptsize $\lambda$}}^j
\nonumber \\
A_{\rm c} &=& \frac{\delta u}{2 h} 
   \sum_{\mbox{\boldmath \scriptsize $\lambda$}} \sum_{jkl}
\left[ \hat E_{{\mbox{\boldmath \scriptsize $\lambda$}}+{\bf e}_k}^{l^{-1}}  -
       \hat E_{{\mbox{\boldmath \scriptsize $\lambda$}}-
      {\bf e}_k}^{l^{-1}} \right]
\textstyle{\prod_{jl}} \hat E_{\mbox{\boldmath \scriptsize $\lambda$}}^j 
  \hat N_{\mbox{\boldmath \scriptsize $\lambda$}}^j 
   \hat N_{\mbox{\boldmath \scriptsize $\lambda$}}^k
\nonumber \\
A_{\rm f} &=&  \frac{1}{h} \sum_{\mbox{\boldmath \scriptsize $\lambda$}} \sum_k
              f_{\mbox{\boldmath \scriptsize $\lambda$}}^k 
     \left[ \hat E_{\mbox{\boldmath \scriptsize $\lambda$}}^{k^{-1}}-
            \hat I \right]
\label{3}
\end{eqnarray}
Here $h$ is the lattice spacing, and $\delta u$ is the 
contribution of one velocity particle to the velocity.
The sum is over lattice sites for 
${\mbox{\boldmath $\lambda$}}$ and over the three dimensions
for $jkl$.  The vector ${\bf e}_k$ gives the lattice site displaced by one unit
in the $k^{\rm th}$ direction.
Expanding this master equation for small lattice spacing, one finds that
it reproduces the Navier-Stokes equation in this limit.

We map this master equation onto a field theory
using the coherent states representation \cite{Peliti,Lee1}.
Negative velocities are treated as
anti-particles.  In this mapping, we find the correspondences
$b_i^*({\bf r}) \leftrightarrow 
\hat E_{\mbox{\boldmath \scriptsize $\lambda$}}^{i^{-1}} $,
$b_i({\bf r}) \leftrightarrow 
   {\hat E_{\mbox{\boldmath \scriptsize $\lambda$}}^i} 
                 \hat N_{\mbox{\boldmath \scriptsize $\lambda$}}^i / h^d$, and
$b_i^*({\bf r}) b_i({\bf r}) \leftrightarrow 
    \hat N_{\mbox{\boldmath \scriptsize $\lambda$}}^i / h^d$.
As is typical, we set $b_i^*({\bf r}) =  \bar b_i({\bf r}) + 1$.
 We find the
following action:
\begin{eqnarray}
S &=& \int_k \int d t~ \hat { \bar b}_i(-{\bf k},t) 
           [ \partial_t + \nu k^2 + \delta(t) ]
                         \hat b_i ({\bf k},t)
\nonumber \\
&& - i \lambda_1 \int_{{\bf k}_1 {\bf k}_2 {\bf k}_3} \int d t~
                  (2 \pi)^d \delta({\bf k}_1 + {\bf k}_2 + {\bf k}_3)
\nonumber \\ 
&&~~~~~~~~~~~ \times
                 k_{2_j} \hat{\bar b}_k^\perp ({\bf k}_1,t)
                          \hat b_k^\perp ({\bf k}_2,t) 
                              \hat b_j^\perp({\bf k}_3,t)
\nonumber \\
&& - i \lambda_2 \int_{{\bf k}_1 {\bf k}_2 {\bf k}_3 {\bf k}_4} \int d t~
              (2 \pi)^d \delta({\bf k}_1 + {\bf k}_2 + {\bf k}_3 + {\bf k}_4)
\nonumber \\ 
&&~~~~~~~~~~~ \times
                 k_{2_j} \hat{\bar b}_k^\perp ({\bf k}_1,t)
                          \hat b_k^\perp ({\bf k}_2,t)
              \hat{\bar b}_j^\perp ({\bf k}_3,t)   \hat b_j^\perp({\bf k}_4,t)
\nonumber \\
&&- \int dt  f_i \hat {\bar b}_i^\perp ({\bf 0},t)
\label{4}
\end{eqnarray}
Here the notation $\int_{\bf k}$ stands for $\int d {\bf k} / (2 \pi)^d$, the
integrals over time are from $t=0$ to some large time $t=t_f$, and
the summation convention is implied.
This action is written in terms of the divergence-free part of the
velocity, $\hat b_i^\perp({\bf k}) = \hat \Pi_{ik}({\bf k}) \hat b_k({\bf k})$. 
For convenience in later calculations, we have included a 
curl-free component in the quadratic terms (Feynman gauge).
Initially, $\lambda_i=1$.

This field theory for fluid mechanics differs from the traditional
one \cite{Wyld,Martin} in that random fluctuations of the fluid are
incorporated by the presence of the $\lambda_2$ term.
Using this action, standard results from fluid mechanics are
reproduced.  For example, a calculation of the
long-time tails in the velocity-velocity
correlation function in two-dimensions is in accord with the
known result \cite{Forster}:
\begin{equation}
\langle v_i({\bf x}, t)  v_j ({\bf x}, t') \rangle \sim
\frac{k_{\rm B} T}{2 \rho} \frac{\delta_{ij}}{4 \pi \nu \vert t-t'\vert
 \ln^{1/2}(\vert t-t'\vert/t_0)}
\label{4a}
\end{equation}
The
scaling of the Kolmogorov energy cascade and the Richardson
separation law are also reproduced
for the common statistical model of turbulence \cite{Orszag2,Majda2}:
\begin{eqnarray}
E(k) &\sim& ({\rm const}) k^{-5/3}
\nonumber \\
r^2(t) &\sim& ({\rm const}) t^3
\label{4b}
\end{eqnarray}
 In both of these calculations, the
$\lambda_2$ term does not contribute.  This is in contrast
to recent work on reaction-diffusion systems, where
fluctuations contribute at low density \cite{Peliti,Lee1,Deem1}.
Since the fluid density is always finite in the present
case, the density fluctuations captured by $\lambda_2$
prove to be irrelevant.

With these preliminaries taken care of, we now turn to the
problem of flow through porous media.  The main effect of the
porous media is to exclude the fluid from certain fixed obstacles.
The velocity is zero within and on the surface of these obstacles.
Within the context of the master equation, if
an obstacle is at 
site ${\mbox{\boldmath $\lambda$}}$, the velocity there must be zero,
$N_{\mbox{\boldmath \scriptsize $\lambda$}}^i = 0$. 
This fixes the lattice spacing as $h \approx R$, where $R$ is the
characteristic size of the blocking particles.
A relatively singular model for the obstacles sets the average 
fluid velocity at the obstacle sites to zero.
With this model, the obstacles generate
the following additional term in the action:
\begin{equation}
e^{-S'} = \prod_{\mbox{\boldmath \scriptsize $\lambda$}} 
 \sum_{n_{\mbox{\boldmath \scriptsize $\lambda$}}=0}^1
   P[n_{\mbox{\boldmath \scriptsize $\lambda$}}] \prod_i 
             \delta \left[ n_{\mbox{\boldmath \scriptsize $\lambda$}} \int dt ~
       \bar b_{\mbox{\boldmath \scriptsize $\lambda$}}^i 
     b_{\mbox{\boldmath \scriptsize $\lambda$}}^i \right]
\label{5}
\end{equation}
here $P[n_{\mbox{\boldmath \scriptsize $\lambda$}}] = (1-\phi) 
     \delta_{n_{\mbox{\boldmath \scriptsize $\lambda$}},0} + 
     \phi \delta_{n_{\mbox{\boldmath \scriptsize $\lambda$}},1}$,
where $\phi$
 is the probability of a site being occupied by an obstacle,
\emph{i.e.}\ the volume fraction of obstacles.
  This condition is difficult to implement directly.
The condition 
$[\int d t~ 
 \bar b_{\mbox{\boldmath \scriptsize $\lambda$}}^i
     b_{\mbox{\boldmath \scriptsize $\lambda$}}^i
]^2  < a^2$, 
where $a^2 \approx 1/(4 \pi \nu R)^2$ is  a constant, effectively implements
this condition in the long-time limit and is much more convenient to 
implement.  That is, we use a Gaussian to regularize the delta function
\begin{eqnarray}
&&e^{-S'} = ({\rm const} )\prod_{\mbox{\boldmath \scriptsize $\lambda$}} 
         \bigg[ (1 - \phi)
+ \phi \prod_i 
\nonumber \\
&&\times      \exp \bigg(- \frac{1}{2 a^2}
           \int d t d t'~ \bar b_{\mbox{\boldmath \scriptsize $\lambda$}}^i(t) 
                        b_{\mbox{\boldmath \scriptsize $\lambda$}}^i(t) 
                         \bar b_{\mbox{\boldmath \scriptsize $\lambda$}}^i(t')
                      b_{\mbox{\boldmath \scriptsize $\lambda$}}^i(t') \bigg) 
           \bigg] 
\label{6}
\end{eqnarray}
Expanding to the lowest relevant order in the fields,
 we find the additional term in the action to be
\begin{eqnarray}
S' &=& ({\rm const}) + \frac{\phi}{2 a^2 h^d} \int d^d {\bf x} \int dt dt'
\nonumber \\ 
&& ~~~~~~~~~~ \times
                           \bar b_i({\bf x},t) b_i({\bf x},t)
                           \bar b_i({\bf x},t') b_i({\bf x},t')
\label{7}
\end{eqnarray}
We set $\gamma_1 = \phi / (2 a^2 h^d)$ for later convenience.
We note that this form looks like a contribution arising from
random, imaginary, velocity-dependent point forces acting on the fluid.

A key aspect of the flow through porous media is the net fluid flow.
To accommodate this, we shift the fields by the average values.
We consider the flow to be along the $x$ direction, and so set
$b_i = \delta_{i,x} v_x + b_i'$ and then rewrite the
action in terms of $b_i'$. Suppressing the primes, we find
\begin{eqnarray}
S &=& \int_k \int d t~ \hat { \bar b}_i(-{\bf k},t)
                [ \partial_t + \nu k^2 - i k_x v_x + \delta(t) ]
                         \hat b_i ({\bf k},t)
\nonumber \\
&&+ i \lambda_1 \int_{{\bf k}_1 {\bf k}_2 {\bf k}_3} \int d t~
                  (2 \pi)^d \delta({\bf k}_1 + {\bf k}_2 + {\bf k}_3)
\nonumber \\ &&~~~~~~~~~~\times
                 k_{1_j} \hat{\bar b}_k^\perp ({\bf k}_1,t)
                          \hat b_k^\perp ({\bf k}_2,t)
                        \hat b_j^\perp({\bf k}_3,t)
\nonumber \\
&&+ \gamma_1 \int_{{\bf k}_1 {\bf k}_2 {\bf k}_3 {\bf k}_4} \int d t d t' ~
                  (2 \pi)^d ({\bf k}_1 + {\bf k}_2 + {\bf k}_3 + {\bf k}_4 )
\nonumber \\ &&~~~~~~~~~~\times
             \hat {\bar b}_i^\perp ({\bf k}_1, t) \hat b_i^\perp ({\bf k}_2, t)
          \hat {\bar b}_i^\perp ({\bf k}_3, t') \hat b_i^\perp ({\bf k}_4, t')
\nonumber \\
&&+ \gamma_2 \int_{{\bf k}_1 {\bf k}_2 {\bf k}_3 } \int d t d t' ~
                  (2 \pi)^d ({\bf k}_1 + {\bf k}_2 + {\bf k}_3 )
\nonumber \\ &&~~~~~~~~~~\times
            \hat {\bar b}_x^\perp ({\bf k}_1, t) \hat b_x^\perp ({\bf k}_2, t)
                \hat {\bar b}_x^\perp ({\bf k}_3, t')
\nonumber \\
&&+ \gamma_3 \int_{\bf k} \int d t d t'  ~
                      \hat{\bar b}_x(-{\bf k},t)
                      \hat{\bar b}_x({\bf k},t')
- \int dt~ f_i \hat{\bar b}_i^\perp ({\bf 0},t)
\nonumber \\
\label{8}
\end{eqnarray}
where $\gamma_2 = 2 v_x \gamma_1$ and $\gamma_3 = v_x^2 \gamma_1$.
We have eliminated the $\lambda_2$ terms, as they do not contribute
in the renormalization procedure.  To incorporate the random
obstacles, we have used the replica trick \cite{Kravtsov1},
but have suppressed these details that do not enter in a
one-loop calculation.

We now apply the renormalization group procedure to this action.
From power counting, the upper critical dimension for this theory 
is $d_c = 3$.  The frictional force on the fluid due to the
obstacles will generate a mass term, however, that eventually renders the
theory finite in any dimension.
We use the momentum shell procedure, where fields on a shell of
differential width $dl$ are integrated out, $\Lambda e^{-d l} < k < \Lambda$,
where $\Lambda \approx \pi / h$ is the cutoff, and $l$ is the 
flow parameter.  
As usual, we rescale time by the dynamical exponent
$t' = t e^{-z dl}$, distance perpendicular to the flow
direction by ${\bf k}_\perp' = {\bf k}_\perp e^{dl}$, and distance along
the flow direction
by the dilation exponent $k_x' = k_x e^{\eta d l}$ \cite{Nelson1998}.
We make use of the average
$\langle \hat{ \bar b}_i^\perp ({\bf k}_1) \hat b_j^\perp ({\bf k}_2) \rangle
= (2 \pi)^d \delta({\bf k}_1 + {\bf k}_2) 
       \hat \Pi_{ij}({\bf k}_2) \hat G_0({\bf k}_2)$
where
$\hat G_0({\bf k}) = 1 / (\nu k^2 - i k_x v_x)$.  
We perform a one-loop calculation, valid to first order in disorder
strength and all orders in the parameter $\lambda_1$.
A large number of integrals, over one-hundred, appear in this renormalization
procedure, and details will be presented elsewhere.
%\cite{todo}.
A typical integral would be one such as
\begin{eqnarray}
\int_{\Lambda e^{-d l} < k < \Lambda} && \hat G_0({\bf k}) \hat G_0( -{\bf k})
\nonumber \\
&=& \frac{\Lambda dl}{(2 \pi)^3} \int_{k \rm{~on~shell}}
 \frac{1}{\nu_\perp^2 k^4 + v_x^2 k_x^2 }
\nonumber \\
&\sim& \frac{\Lambda dl}{(2 \pi)^3} \int_{k_\perp {\rm ~on~shell}} 
                \int_{-\infty}^\infty
               \frac{d k_x}{\nu_\perp^2 k_\perp^4 + v_x^2 k_x^2 }
\nonumber \\
&=& \frac{\Lambda dl}{(2 \pi)^3} \int_{k_\perp {\rm ~on~shell}}
            \frac{\pi}{\nu_\perp v_x k_\perp^2}
\nonumber \\
&=& \frac{1}{4 \pi v_x \nu_\perp} dl
\label{9}
\end{eqnarray}
We have made use of the fact that the integral over $k_x$ converges, and so
it can be taken off the shell \cite{Nelson1998}, and we have used the fact that
$\nu_x$ flows to zero and can be ignored in these calculations, \emph{i.e.}\ 
we replace $\nu k^2$ by $\nu_\perp k_\perp^2$ in the propagator, where
$k^2 = k_x^2 + k_\perp^2$.

When computing the renormalization flows, some additional terms
are generated.  We present the details only for the $x$-fields, as 
these are most significant.  A term such as
$
 \gamma_{1{\rm u}}
      \int_{{\bf k}_1 {\bf k}_2 {\bf k}_3 {\bf k}_4} \int d t d t' ~
                  (2 \pi)^d ({\bf k}_1 + {\bf k}_2 + {\bf k}_3 + {\bf k}_4 )
            \hat {\bar b}_x^\perp ({\bf k}_1, t) \hat b_x^\perp ({\bf k}_2, t)
          \hat {\bar b}_x^\perp ({\bf k}_3, t') \hat b_x^\perp ({\bf k}_4, t)
$
is generated.  In addition, a mass term is generated,
$
m_x \int_{\bf k} \int d t~ \hat { \bar b}_x(-{\bf k},t) 
                         \hat b_x ({\bf k},t)
$.
Defining the dimensionless couplings
$g_x = \gamma_1 /(4 \pi \nu_\perp v_x)$, 
$g_{x{\rm u}} = \gamma_{1{\rm u}} /(4 \pi \nu_\perp v_x)$, 
$g_2 = \gamma_2 / (8 \pi \nu_\perp v_x^2 )$, and
$g_3 = \gamma_3 / (4 \pi \nu_\perp v_x^3) $, we find the
following flow equations:
\begin{eqnarray}
\frac{d \ln  g_3}{dl} &=& (2 z - d + 1 - \eta) 
       - 2  g_x - 2   g_{x {\rm u}}
\nonumber \\
\frac{d \ln  g_2}{dl} &=& (2 z - d + 1 - \eta) 
        - 2  g_x - 2  g_{x {\rm u}} 
\nonumber \\ &&
                 +      \lambda_{1x} (g_x +g_{x {\rm u}}) g_3/g_2
\nonumber \\
\frac{d \ln g_x}{dl} &=& (2 z - d + 1 - \eta) 
       - 2 g_x + 2 \lambda_{1x}   g_2 -
                         \lambda_{1x}^2  g_3
\nonumber \\
\frac{d \ln g_{x {\rm u}}}{dl} &=& (2 z - d + 1 - \eta)
       - 2 g_{x {\rm u}}   -
     4 g_x 
 + 4 \lambda_{1x} g_2 
\nonumber \\ &&
      + 2 \lambda_{1x}  g_2 g_x/g_{x {\rm u}}
              -\lambda_{1x}^2  g_3 g_x/g_{x {\rm u}}  - 2 \lambda_{1x}^2 g_3
\nonumber \\
\frac{d \ln \lambda_{1x}}{dl} &=&  (z - \eta)
             -2  g_x   - 2 g_{x {\rm u}} + 
                        4 \lambda_{1x} g_2
              - 2 \lambda_{1x}^2 g_3 
\nonumber \\
\frac{d m_x }{dl} &=& z m_x + 
             2 g_x \nu_\perp \Lambda^2  
                   + 2 g_{x {\rm u}} \nu_\perp \Lambda^2
- 2 \lambda_{1x} g_2 \nu_\perp \Lambda^2 
\nonumber \\ 
\frac{d \ln v_x }{dl} &=& 
     (z - \eta)  -4 \lambda_{1x} g_2  + 3 \lambda_{1x}^2 g_3
,~~~~~
\frac{d \ln f}{d l} = z
\nonumber \\ 
\frac{d \ln \nu_\perp }{dl} &=& 
    (z - 2) - 2 \lambda_{1x} g_2  - 
                \frac{1}{2} \lambda_{1x}^2 g_3
\label{10}
\end{eqnarray}
At long times, we find 
$g_x(l) \sim g_2(l) \sim g_3(l)  \sim 1/(g_0^{-1} + 2 l)$.
We find the unphysical term $g_{x {\rm u}} (l)$ should vanish.  We also find
$\lambda_{1x} \sim \lambda_1^0 / (1 + 2 l g_0)$.  From the flow equation for
the viscosity, we find the dynamical exponent is $z = 2 + O(l^{-2})$.
From the flow equation for the velocity, we find the dilation
exponent is $\eta = 2 + O(l^{-2})$.  
The mass quickly
reaches the asymptotic form
$m_x(l) \sim m^* e^{2l}$ whatever the
non-universal, $\Lambda$-dependent terms in $dm_x / dl$.
In the absence of the mass term, \emph{e.g.}\ for imaginary,
velocity-dependent
forces with zero average value, these flow equations generate
interesting scaling behavior.  In the present case, the generated mass
stops the flow at a characteristic value of the flow  parameter, $l^*$.

We determine the value of the generated mass by 
a force balance argument.
 From the requirement that
$\langle b_x \rangle = 0$, we find $m_x(l) v_x = f(l)$, and
conclude that $f_0 = m^* v_x$.  From fluid mechanics, 
it is known that the force density is given by Stokes' Law
at low volume fraction of obstacles and moderate flow rates \cite{Brady}:
$f_0 = 6 \pi R \nu_\perp v_x \phi_0 / (4 \pi R^3/3)$, where $R$ is the
radius of the spherical obstacles.  Equating these two results,
we find $m^*/\nu_\perp = 9 \phi_0 / (2 R^2)$.  We further
calculate the correlation length in the fluid
$\xi_x = e^{2l} \xi_x(l) = e^{2l} [\nu_x(l) / m_x(l)]^{1/2}
= e^{2l}[\nu_0 e^{-2l} / (m^* e^{2l})]^{1/2}$:
\begin{equation}
\xi_x = \left(\frac{2 R^2}{9 \phi_0} \right)^{1/2}
\label{11}
\end{equation}
 which is exactly the Brinkman screening length \cite{Brady}.
At high flow rates, empirical expressions for the drag force
are available that can be used to identify $m^*$ \cite{Bird}.

We now address the issue of the dispersion coefficient.
The dispersion coefficient $D_x = \int_0^\infty dt' 
[\langle v_x({\bf x},t) v_x({\bf x}+\hat {\bf x} v_x t', t+t') \rangle - 
  v_x^2]$
is a measure of the ``dispersive'' transport induced by the
fluid.  
We can calculate this quantity directly from the
field theory by perturbation theory.
We find $D_x(l) = g_3(l) v_x^2 / m(l)$.
Using the approximate value of $a^2$ and the determined form
of $m(l)$, we find $D_x(l) = e^{-2 l} 4 \pi R v_x / 9 $.
We match the flow equations to this perturbation theory
when the flow equations
begin to loose their validity, $m_x(l^*) \approx \nu_\perp \Lambda^2$, which
gives the characteristic flow parameter $l^* = \frac{1}{2}
\ln (\nu_\perp \Lambda^2/m^*)$.
  We calculate the dispersion coefficient from matching \cite{Deem1} as
$D_x^* = e^{2 l^*} D_x(l^*)$.  We find the renormalized dispersion
coefficient to be
\begin{equation}
D_x^*  = \frac{4 \pi}{9} R v_x
\label{12}
\end{equation}
This  dispersion coefficient
is due solely to mechanical dispersion of
the fluid, and the form is in exact agreement with
the mean-field result of Koch and Brady \cite{Brady}.
The dispersion coefficient is proportional to the
P{\'e}clet number and independent of the volume fraction in the limit
of high P{\'e}clet number and low volume fraction.
Only the numerical prefactor, here given approximately, is non-universal.
Interestingly, the asymptotic value of the dispersion coefficient
is independent of $l^*$, as long as the mass has been generated
and $l^* g_0 << 1$, which is the
case for small volume fractions or large
P{\'e}clet numbers.  The independence of the
observable on degree of renormalization is common when mean-field theory is
essentially exact, as is the case here.

\begin{figure}[tbp]
 \centering
 \leavevmode
\psfig{file=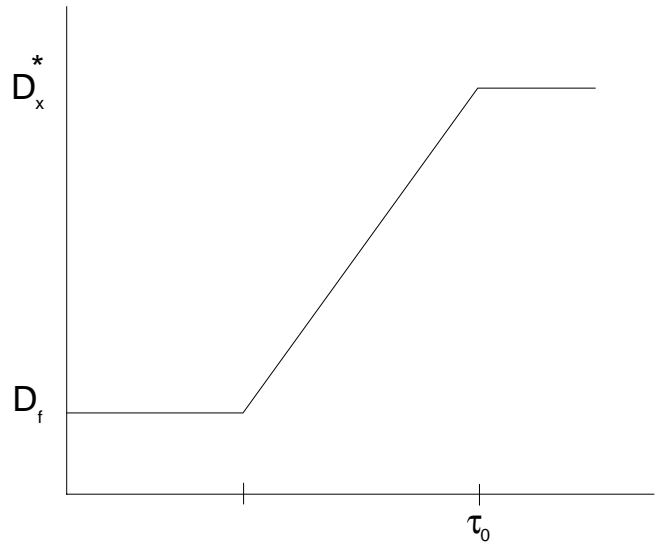,width=2.8in,angle=-90}
\caption
{
Shown is the behavior of the dispersion coefficient at low
volume fraction of disorder.  The dispersion coefficient 
reaches its asymptotic value at a dimensionless time of
$\tau_0 = Pe^{1/3}$.
}
\label{fig1}
\end{figure}

In the dilute limit, the dispersion coefficient grows as
shown in figure \ref{fig1}.  The dispersion coefficient reaches
its asymptotic value at a time on the order of $t_0$.  If time is
nondimensionalized as $\tau = t v_x/R$, this characteristic
time is given by $\tau_0 = Pe^{1/3}$ \cite{Brady}, which
is the time scale for diffusive transport across
the boundary layer.  

The suggestion that dispersion coefficients diverge logarithmically
in three dimensions at large P{\'e}clet numbers and finite
volume fractions of disorder \cite{Frenkel}, then, is not borne
out by the present calculations.  
Such a divergence would show up as a $l^*$ dependence of the
predicted dispersion coefficient, which is absent in our
calculations.
The proposal of a diverging dispersion coefficient
was based upon simulation
data, and these simulation data may not have sampled the full
diffusive boundary layer transport that occurs on a time scale
that grows as $Pe^{1/3}$.  While transport at high volume fractions
is, in principle, not accessible by our renormalization group
calculations, physical aspects of the expected behavior are present
in our result.  
In particular, curing of the logarithmic divergence in the theory by
growth of a mass term due to the frictional drag force exerted by the
disorder particles on the fluid seems generic.  
This frictional
drag force would seem to prevent divergences at any volume fraction of
disorder.  Indeed, the higher the volume fraction of disorder, the
greater the expected accuracy of the mean-field result.
Our result is in agreement with the mean-field theory  as soon
as the frictional mass has been generated.

The renormalization group treatment of dispersion for a particle-based
model shows that mean-field theory misses no essential physics at low
volume fractions.

%This research was supported by the National Science Foundation
%through grant no.\ CTS--9702403.

\hbox{}
\vspace{-0.35in}
\bibliography{fluid}

\end{document}